# Fabrication of multilayer edge molecular electronics and spintronics devices

Pawan Tyagi,

Department of Chemical and Materials Engineering, University of Kentucky, Lexington, Kentucky-40506, USA

Current Address: School of Engineering and Applied Science, University of the District of Columbia, Washington DC-20008, USA

**Abstract:** Advancement of molecular devices will critically depend on the approach to establish electrical connections to the functional molecule(s). We produced a molecular device strategy which is based on chemically attaching of molecules between the two magnetic/nonmagnetic metallic electrodes along the multilayer edge(s) of a prefabricated tunnel junction. Here, we present the fabrication methodology for producing these multilayer edge molecular electronics/spintronics devices (MEMEDs/MEMSDs) and details of the associated challenges and their solutions. The key highlight of our MEMED/MEMSD approach is the method of producing exposed side edge(s) of a tunnel junction for hosting molecular conduction channels by *a simple liftoff method*. The liftoff method ensured that along the tunnel junction edges, the minimum gap between the two metal electrodes equaled the thickness of the tunnel barrier. All of the tunnel junction test beds used a ~2 nm alumina (AlOx) tunnel barrier. We successfully bridged the magnetic organometallic molecular clusters and non-magnetic alkane molecules across the AlOx insulator along the exposed edges, to transform the prefabricated tunnel junction into the molecular electronics or spintronics devices. Tunnel junction test beds were fabricated with a variety of metal electrodes, such as NiFe, Co, Ni, Au, Ta, Cu and Si. Stability of ultrathin thin AlOx varied with the type of bottom metal electrodes used for making MEMED/MEMSD. Additionally, molecular solution used for bridging molecular channels in a MEMED was not compatible with all the metal electrodes; molecular solution resistant ferromagnetic electrodes were developed for the fabrication of MEMSDs. MEMSD approach offers an open platform to test virtually any combination of magnetic electrodes and magnetic molecules, including single molecular magnets.

## 1. Introduction:

Ultimate electronics device miniaturization is possible by the utilization of a discrete molecule as a memory, logic or switching element in the futuristic devices. A benefit to this scheme is that the few angstrom long molecules can be synthesized reproducibly with well-defined electronic states tuned through chemical functionality. Molecular conductance has been investigated via a variety of theoretical [1] and experimental methods, and several interesting device characteristics were observed. Molecular devices have exhibited useful phenomenon like negative differential resistance [2], spin-state dependent conduction [3, 4] and stable memory function based on conformational shifts [5]. However, until now none of the fabrication approach has allowed the facile integration of molecules to function in the CMOS type circuit architecture. For molecular devices to become a technological reality, stable metallic electrodes must be engineered to stay at a nanometer scale robust gap to allow the incorporation of bistable molecules. It must be noted that a molecule should not be used as a physical spacer between two metal electrodes; such device will neither be mechanically stable nor the transport through the molecule will be steady due to everlasting structural changes.

Can thin film thickness be used as a 'molecular dimension spacer' between two metallic electrodes? Can metallic electrodes present on either side of the spacer be chemically bonded with the molecule(s) of interest to produce the molecular electronics devices? Recent developments in molecular device fabrication provide affirmative answer to these questions [6]. Conventional thin film deposition processes can control the film thickness with angstrom-level precision. On the edges of a crisply cut multilayer thin-film structure, there are by default parallel lines of material that have widths corresponding to the thickness of deposited films. Exposed side edges of thin films have been successfully utilized as a template for growing nanostructures [7]. However, utilizing the edges of thin films into nanoscale devices involve integration challenges, and required judicious consideration about the potential role of individual films of a multilayer to be used in a device. Here we describe the approach in which the sides of a tunnel junction are used for producing molecular electronics and spintronics devices. Tunnel junction configuration is extremely useful in the sense that an ultrathin insulator work as a molecular dimension spacer between two arbitrarily chosen metallic electrodes. Along the tunnel junction edges the metallic electrodes can be chemically attached to the molecular bridges. The molecular bridges on the tunnel junction's edge can eventually over take the transport via the insulator/spacer in the planar area. These multilayer edge molecular electronics devices (MEMEDs), which employ a tunnel junction edge(s) produce a new approach of integrating molecule as a device element. Several groups have utilized MEMED approach for producing molecular electrodes. Here we first review the existing molecular electrodes approaches. Then we discuss our liftoff based MEMED (liftoff-MEMED) fabrication scheme, which was utilized for producing MEMEDs.

## 1.1 Review of multilayer edge molecular electronics devices:

A number of research groups [8-11] have produced various configurations of tunnel junctions to realize MEMEDs. Following is the review of these MEMEDs. Here, MEMEDs' classification is based on the methods of producing metal-insulator-metal junctions with the exposed side edges.

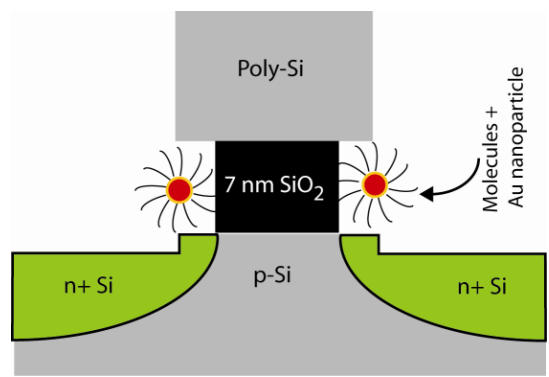

Fig.1: Nano-gap between n$^+$ Si and poly silicon (p-Si) electrodes. Nano-gap was bridged by the molecule and gold nanoparticle assembly. Schematic design adopted from ref. [7]

### 1.1.1 Semiconductor-insulator-semiconductor junction with etching produced exposed side edges:
The first MEMED was realized with semiconductor-insulator-semiconductor junction [8] (Fig. 1a). Conventional CMOS fabrication steps were performed to produce a ~7 nm thick silicon dioxide (SiO$_2$) between two conducting silicon (Si) electrodes. The SiO$_2$ insulating layer was chemically etched for producing exposed side edges for hosting molecular channels. Along the exposed sides the minimum gap between two silicon electrodes was equal to SiO$_2$ thickness (Fig. 1a). The ~ 7 nm thickness of SiO$_2$ was significantly more than the physical length of the individual molecules used in this study. Hence, a multi-step chemistry was performed to chemically bond the molecules with nanoparticles to produce > 7 nm long hybrid conduction channels. Here, a conduction bridge was composed of short molecules with diazonium functional group, and a 5 nm gold particle. Current-voltage (I-V) studies revealed that hybrid conduction channels increased the current by several orders over the background level. This study clearly established that overlap of metallic wave functions via conduction channels, along the junction edge, is much stronger than that via planar SiO$_2$ tunnel barrier.

An improved version of semiconductor-insulator-semiconductor based MEMED was produced by metalizing the exposed side edges [12]. Gold (Au) metalized exposed sides of aluminum gallium arsenide (AlGaAs)/~8 nm gallium arsenide (GaAs)/ AlGaAs facilitated the chemical bonding of molecular bridges. The 8.5 nm long oligo-(p-phenylenevinylene) (OPV) molecules were successfully bridged across a GaAs spacer. According to I-V studies at 4.2 K, OPV molecules significantly increased the charge transport rate.  However, these schemes have limited scope for further advancement of MEMED. A ~7 nm insulator preclude the use of many exotic molecules with 1-3 nm length. Additionally, the surface chemistry steps for the attachment of molecular bridges are significantly complex, and can be affected by ongoing semiconductor oxidation. Usage of various metallic electrodes, especially ferromagnetic metals, is also not straight forward. Transforming these approaches for producing molecular spintronics devices will necessitate the incorporation of ferromagnetic electrodes.

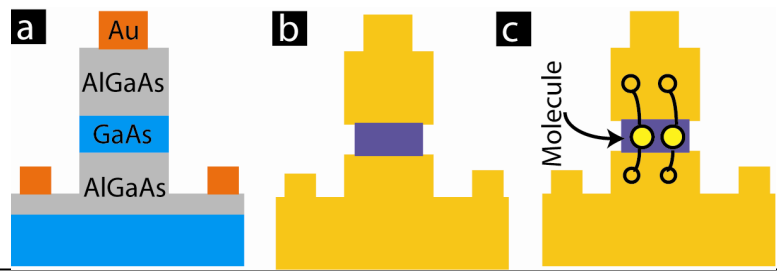

Fig. 2: Metalized semiconductor films stack with exposed side edges: (a) Cleaving produced AlGaAs/GaAs/AlGaAs exposed edges. (b) Cleaved edges were metalized with gold for, (c) easy attachment of molecular channels. Panels (a-c) adopted from ref. [9]

**1.1.2 Au-SiO$_2$/SiNx-Au junction with chemical etching produced exposed side edges:**

Further improvements for the MEMED fabrication were made by (a) reducing the insulator thickness and (b) incorporating metal electrodes in a simpler manner. Ashwell et al.[10] produced tunnel junctions with gold (Au) metal electrodes and with 3.5 nm insulating thickness at the junction's edges (Fig. 3). A mushroom shaped tunnel barrier was produced for realizing the thicker tunnel barrier, away from the exposed side edges, to reduce the leakage current. Tunneling barrier is made up of two insulators: silicon nitride (SiNx) and SiO$_2$. The SiO$_2$'s dome shaped profile was realized by using an undercut profile in the photoresist pattern, through which SiO$_2$ was deposited. However, SiNx under SiO$_2$ was etched to produce a mushroom shaped tunnel junction edge profile. Etching step set the minimum gap between two gold metal electrodes along the edges. Bridging the molecular channels across the vertical gap increased the effective device current by >3 orders over the background current. It is noteworthy that the molecular channels were produced in two steps. In the first step 4-[(3-mercaptophenylimino)-methyl]-benzaldehyde anchors were chemisorbed on to the two gold electrodes. In the second step 2,6-diaminoanthra-9,10-quninoe molecules were bridged between the two anchor molecules to produce complete conduction channels. Ashwell et al.[10] successfully performed a reversible experiment to disconnect the molecular conduction channels between two metal electrodes to retrieve the background leakage current via mushroom shaped tunnel barrier. In this experiment, molecular channels were curtailed to undo the effect of molecules; device current reverted back to the background level. This crucial reversible experiment substantiated the fact that only molecular channels increased the current over the background current.

The major limitation of this study is the complexity in growing a mushroom shaped bi-layer insulator. Setting an insulator thickness <2 nm is also significantly challenging. Establishing a ~2 nm insulator thickness is essential to bridge the molecular channels directly and without the need of additional anchoring molecules. In this study only gold metal electrodes were utilized. For the realization of molecular spintronics devices ferromagnetic electrodes are need to be employed.

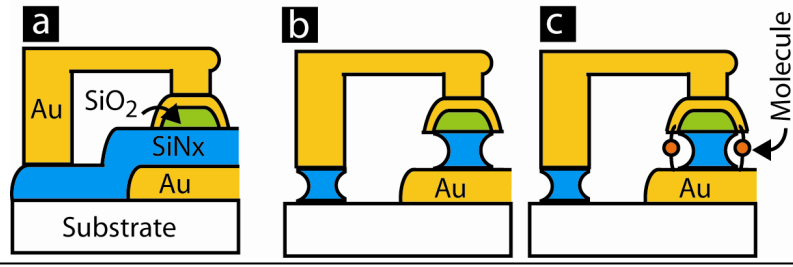

Fig. 3: Ashwell et.al.'s MEMED. (a) Au/ SiO$_2$/SiNx/Au tunnel junction with unexposed sides. (b) Chemical etching receded SiNx under top Au electrode to set the minimum inter-electrode along the edges. (c) bridging of molecular channels across the insulator gap. Adopted from ref.[9].

### 1.1.3 Pt-AlOx-Pt with chemical etching produced exposed side edges:

Chen et al.[9] produced a simpler version of MEMED by utilizing ~3 nm alumina (AlOx) and platinum (Pt) electrodes. Resulting Pt-AlOx-Pt tunnel junction was chemically etched to remove the AlOx from the planar area. Etching step produced exposed side edges for the creation of molecular conduction channels across ~3 nm AlOx. In this study AlOx insulator was grown by the oxidation of a pre-deposited aluminum (Al) film (Fig. 4). A low leakage current was achieved by reducing the roughness of the bottom electrode from 0.6 nm to 0.3 nm. For electrode smoothing an energy intensive Ar ion sputtering was performed. It is noteworthy that a smoother bottom electrode can be covered with the thinner AlOx, yet producing a low leakage current. For further reduction in background current, the tunnel junction area was kept to be of nanoscale by employing E-beam lithography. Typical junction area was ~0.04 $\mu m^2$. *Other*

*MEMED* discussed here possessed tens of μm² junction area. In the present case, oligo-phenylene-ethylene (OPE) molecules terminated with isocyanide groups were utilized as the molecular conduction channels. Isocyanide group enabled metal-molecule bonding to produce molecular conduction bridges across the AlOx insulator (Fig. 4d). Molecular channels were shown to enhance the current by more than one order. This study did not show any reversible experiment where molecular channels were destroyed to retrieve the tunnel junction's leakage current. This study is a nice demonstration of MEMED with ~3 nm insulator gap. However, more control experiments are required to ensure the durability of this type of MEMED. This approach utilized Pt metal electrodes; ferromagnetic electrodes are to be used as metal electrodes for producing molecular spintronics devices using this approach.

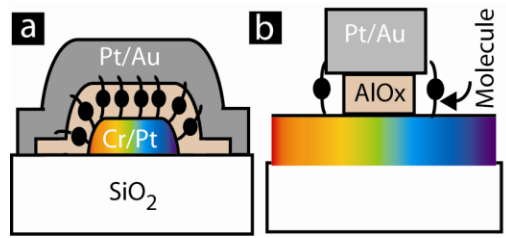

Fig. 4 (a) Attached molecules along the exposed sides of Pt/AlOx/Pt tunnel junction. (b) Cross sectional view of the same tunnel junction ref. [8].

**1.1.4 NiFe-AlOx-NiFe junction with ion milling produced exposed side edges:**

Development of the molecular spin devices need tunnel junction with ferromagnetic electrodes and durable thinner insulator. Ultrathin insulators (< 2 nm) have been employed by the heavily studied magnetic tunnel junctions (MTJ). A MTJ being a tunnel junction with ferromagnetic electrodes [13, 14] can be transformed into a multilayer edge molecular spintronics device (MEMSD). We made several attempts to modify MTJ into molecular devices [15]. The key challenge was to retain insulator only under the top electrode and remove the additional insulator from the planar area. This step was necessary to set the distance between two metallic electrodes along the junction edges to be equal to the thickness of insulator in the planar area. In the initial approach we used ion milling to remove the extra AlOx insulator from the planar region (Fig. 4a-b). This *ion milling* step created exposed side edges for hosting molecular channels (Fig. 4); an Argon ion beam was bombarded to do dry etching of unwanted tunnel junction area [11]. A NiFe soft magnetic layer was used as the ferromagnetic electrodes. NiFe showed excellent stability in the ambient conditions[16]. NiFe remain undamaged by the electrochemical molecule attachment protocol used in our studies[15]. As a key advantage NiFe bottom electrode could be directly sputtered to yield ~0.1-0.2 nm roughness. Generally, a ~2 nm AlOx was produced by the plasma oxidation of a pre-deposited ~1.5 nm Al film. Exposed edge NiFe/AlOx/NiFe junction with ~10 $\mu m^2$ planar area exhibited a nonlinear current-voltage graph, where current was in $\mu A$ range. Organometallic molecular clusters with functionalized alkane tethers were bridged across the tunnel barrier [11]. Creation of the molecular bridges increased the tunnel junction current by ~1 fold. As a major drawback, ion-milled tunnel junctions were unstable and short lived, at least with the experimental conditions used for this study [11]. In a relevant study we found that the process induced tensile and compressive stresses caused time dependent tunnel junction failure [17]. Additionally, due to the unavailability of right ion milling experimental setup this approach was impractical and extremely cumbersome to carry on. New improvements in MEMED approach were incorporated to produce stable and versatile MEMSD using in house experimental capabilities. The new MEMED design is based on liftoff method and discussed in the following section.

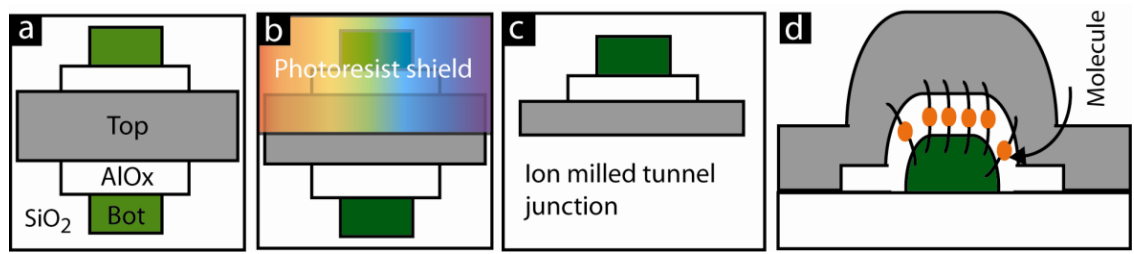

Fig. 5: (a) Top view of as produced metal/AlOx/metal tunnel junction (b) Creation of the photoresist shield to protect a part of tunnel junction from ion milling. (b) Top view of the ion milled tunnel junction. (d) Side view of the exposed edge where molecules are bridged.

Molecular spintronics is the subject of intense research due to their potential role in revolutionizing computational devices [15, 18]. MEMED approaches can yield molecular spintronics devices if a tunnel junction test bed has ferromagnetic electrodes. Molecular spintronics devices based on the MEMED approach can be named as multilayer edge molecular spin devices (MEMSD). Several designs have been used for the experimental demonstration of MEMED approach [8-11]. However, all the MEMED designs are not equally suitable for developing a MEMSD [6]. Two major hurdles in realizing MEMSDs are: (i) prepare tunnel junction test bed with ferromagnetic electrodes, and insulator thickness to be smaller than the length of target molecules, and (ii) prepare exposed side edges in such a way that along the tunnel junction edges the distance between two ferromagnetic electrodes is equal to the insulator film thickness. These two hurdles were surmounted by the ion milling based approach (Fig. 5) to some extent; but, this approach was not feasible due to the lack of resources. More importantly final MEMSDs turned out to be unstable due to stress induced by ion milling step. Here we discuss an improved methodology which utilized simple liftoff step for producing exposed sides along the tunnel junction edges for establishing molecular conduction channels. Liftoff based approach ensured that inter-electrode distance along the tunnel junction edges was equivalent to the insulator film thickness which could be tailored to host the desired molecules [11]. Liftoff based MEMED [15] fabrication scheme was developed to produce a molecular device without incorporating an etching step. A liftoff based MEMED can incorporate virtually any thin film configuration, including that of magnetic tunnel junction (MTJ). Hence, this approach is highly suitable for the development of MEMSD. Remaining part of this paper focuses on the fabrication and characterizations of a stable lift off based MEMED/MEMSD. This paper reports our efforts for optimizing the mechanical and chemical stability of the liftoff based MEMED/MEMSD.

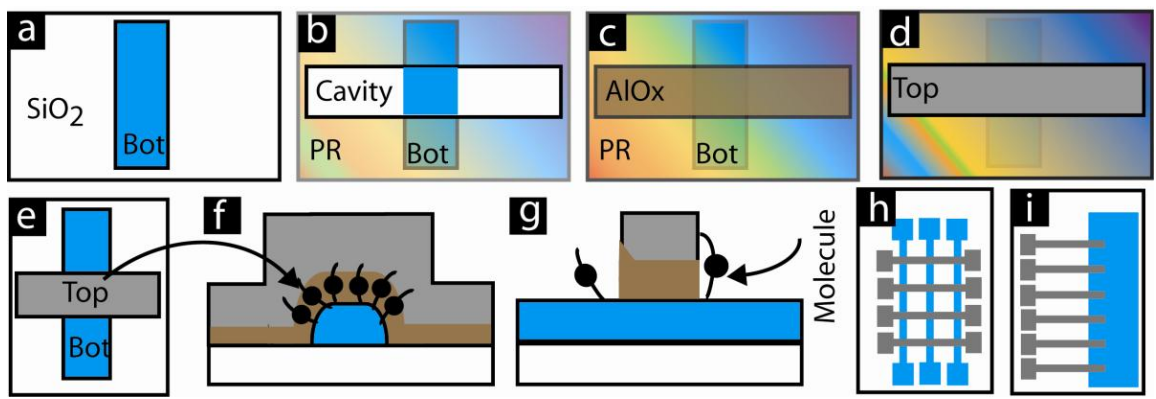

Fig. 6: Liftoff produced MEMED: (a) Bottom electrode on insulating substrate. (b) Cavity in photoresist for the deposition of (c) ~2 nm AlOx and (d) top metal electrode. (e) Top view of exposed edge tunnel junction. (f) Side view of an exposed edge with molecular bridges. (g) Cross sectional view of the tunnel junction, showing molecular attachment is not favored when insulator thickness is increased by unclean liftoff. (h) Cross bar and (i) isolated junction designs used in MEMSD approach.

## 2. Experimental details:

Throughout this work, a silicon wafer with thermally grown 100 nm silicon di oxide ($SiO_2$) isolation layer was used as the substrate. Prior to every photolithography step, substrates were sequentially cleaned with acetone, isopropyl alcohol and de-ionized (DI) water. Samples were then dried in nitrogen jet. The dimensions of the top and bottom electrodes were defined by a conventional photolithography method using Shipley 1813 positive photoresist and a Karl Suss mask aligner. Photolithography step was crucial for developing the optimum bottom electrode's edge profiles; slightly undercut profile was necessary to avoid notches on the metallic pattern after liftoff. The photoresist thickness, exposure time and developing steps were optimized to obtain a suitable edge profile, for the deposition of the first electrode. Metal electrodes were sputter deposited with 100-150 W RF sputtering gun power and 1-2 mtorr argon (Ar) pressure in AJA International Multitarget Sputtering System. After the required film deposition steps, a gentle liftoff process was carried out with a Shipley 1165 resist remover for 2-4 hours (Fig. 6a). For all the depositions, base pressure in sputtering machine was ~2 x $10^{-7}$ torr.

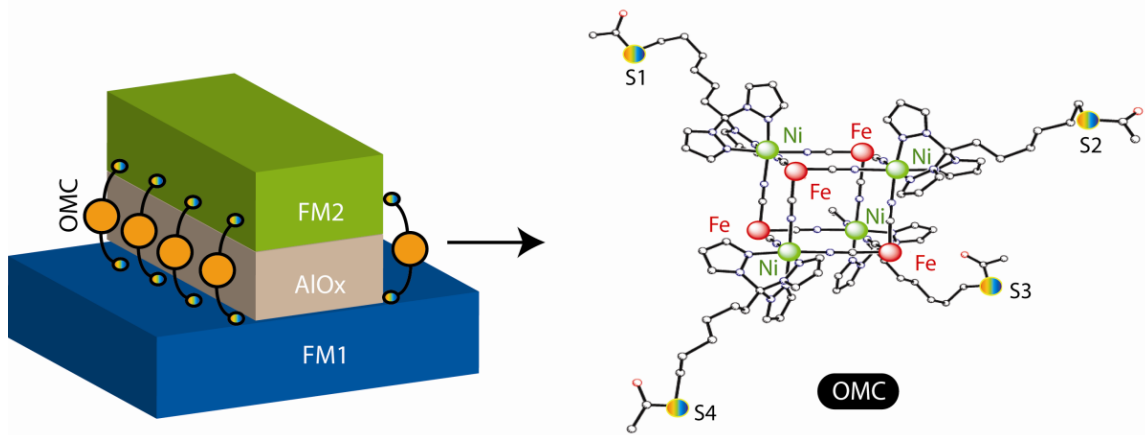

Fig. 7: Schematic of a liftoff produced magnetic tunnel junction transformed into MEMSD. In the right side of schematic is the crystal structure of organometallic molecular cluster (OMC). Cubic cluster of OMC is consisting of alternating $Fe^{III}$ and $Ni^{II}$ centers (corners). Fe and Ni pairs are bridged by cyano groups (edges). All anions, hydrogen atoms, pendant pyrazoles, and disordered S(acetyl)hexyl- chains are not shown for clarity. Note: 1-S(acetyl)tris(pyrazolyl)decane tethers were used in MEMSD.

The key highlight of the liftoff based MEMED (liftoff-MEMED) is that the ultrathin AlOx insulator and the top metal electrodes were deposited through the same photoresist pattern (Fig. 6b) [11, 19]. Utilization of the same photoresist cavity, (Fig. 6b), helped matching the lateral dimensions of the AlOx and the top metal electrode. When lateral dimensions of AlOx and top electrodes are the same then inter-electrode gap along the junction edges is equal to the AlOx thickness (Fig. 6a-c). The second photolithography step was performed with diluted photoresist; diluted photoresist helped reducing the photoresist thickness in 300 to 500 nm range, which was necessary to mitigate the shadow effect from the photoresist sidewall. The AlOx and top metal electrodes were deposited through same photoresist cavity (Fig. 6c-d) RMS roughness measured by AFM for all the metal films was around ~0.2 nm. Deposition of the 2 nm thick AlOx insulator, the most important step of molecular electrode fabrication scheme, was accomplished using a multi-step process. To attain a smooth coating of the insulator, aluminum (Al) metal was deposited on the bottom metal electrode; the similar surface energies of metals result in good wetting without island growth. The fabrication of 2 nm AlOx was accomplished in two steps. Al metal was deposited from a DC powered gun at 75 W power and 2 mtorr Ar pressure. AlOx was deposited in two steps. First 1 nm Al film was deposited and then it was plasma oxidized with 60 mtorr of 1:1 Ar and oxygen gas mixture for 30 seconds at 20 W substrate biases in sputtering chamber. A second 0.5 nm Al deposition was followed by the same plasma oxidation parameters as used for the 1 nm Al in the first step. This double step AlOx fabrication scheme gave an optimal tunnel barrier quality for a ~2.0 nm thick insulating layer. The ~2 nm AlOx tunnel barrier exhibited a breakdown voltage ranging from 1.8 to 2.5 V. Liftoff of photoresist produced an exposed edge tunnel junction (Fig. 6e). To ensure that no photoresist remained on the side walls of the exposed edge tunnel junction, occasionally an additional step was performed. Completed tunnel junctions were exposed in UV light and then samples were submerged in alkaline developer. Generally, this step was performed when expired photoresist was in use.

Subsequently, molecular clusters were bridged across ~2 nm AlOx to establish the molecular-conduction channels (Fig. 6f). Holmes and co-workers developed the molecular complex (Fig. 7b) with functionalized alkane tethers [20]. These alkane tethers with thiol ends, can be bridged across AlOx insulator along MEMED edges (Fig.7a). This organometallic molecular cluster (OMC) have following configuration, $\{[(pzTp)Fe^{III}(CN)_3]_4[Ni^{II}(L)]_4[O_3SCF_3]_4\}$ (pzTp = tetra(pyrazol-1-yl)borate; L = 1-S(acetyl)tris(pyrazol-1-yl)decane) or L = 1-S(acetyl)tris-(pyrazol-1-yl)hexane [20]. In an OMC, the $Fe^{III}$ and $Ni^{II}$ centers resided in alternate corners of a slightly distorted box and are linked via cyanides (Fig. 7). The cluster core contained coordinated tris(pyrazolyl)decyl- chains that are terminated with S(acetyl) groups at alternate corners ($Ni^{II}$ centers) of the cubic cluster [21]. The S(acetyl) groups were chosen for several reasons: (1) acetyl group-protected-thiol groups have been extensively used in self-assembled monolayer (SAM) based molecular electronics devices. (2) acetyl protection prevents sulfur atom coordination to transition metal centers during complexation, (3) acetyl protection enable the use of air sensitive thiols as a functional group in a molecular device, and (4) acetyl groups are easily removed by electrochemical or chemical means [22].

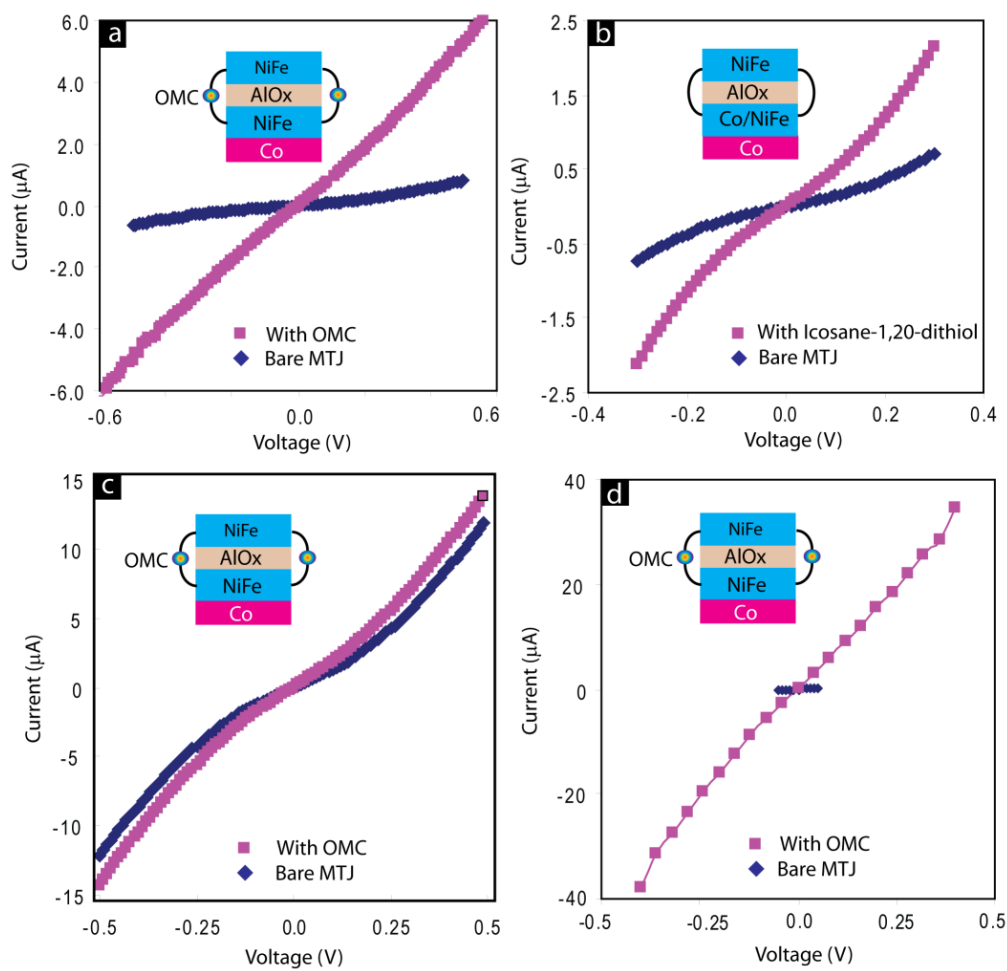

Fig. 8: As produced current-voltage studies on MTJs (Co/NiFe/AlOx/NiFe) before and after molecule attachment: (a) MTJ in cross bar geometry and with organometallic molecular clusters (OMCs), (b) MTJ in cross bar geometry and with Icosane-1,20-dithiol alkane molecules (c) MTJ in cross geometry and with OMC showing relatively small current increase, and (d) MTJ in isolated "T" shaped junction showed same molecule's effect as observed with cross bar geometry. Bare MTJ current-voltage data was recorded in ±0.05 V range. To avoid high bias induced instability only few molecular junctions were subjected to large bias current-voltage study before and after the molecule attachment.

OMCs were bridged across the MEMED's insulator gap in the exposed region via the thiolacetyl termini. The electrochemical attachment was accomplished via immersion of the electrodes in a dichloromethane solution of OMC, followed by the application of alternating a ±100 mV bias between the two metal electrodes. The voltage pulses occurred at a time interval of 0.01 seconds for 2 min. After electrochemical molecule attachment MEMEDs were rinsed with dichlromethane, 2-propanol, and deionized water and then dried under a nitrogen gas stream. Immersion of the MEMEDs in a dichloromethane solution of OMC, without the voltage pulses, did not result in conclusive molecule attachment. We surmised that the use of alternating voltage during electrochemical step helped in conditioning the metal electrode surface by reducing it and subsequently favored the metal-thiol bonding. Alternating voltage also appeared to deprotect thiol group from acetyl group and hence enabling molecule attachment. Tour group experimentally observed that electrochemically directed assembly was significantly faster than the conventional self-assembly of thioacetyl-terminated oligo(phenylene ethynlyene)s on gold and platinum surfaces [22].

A number of molecular electrodes were produced using various thin film configurations: magnetic tunnel junctions (Co/NiFe/AlOx/NiFe), non-magnetic tunnel junction (Ta/AlOx/Ta and Cu/AlOx/Cu), and dissimilar electrodes with tunnel junctions (Nickel (Ni)/AlOx/Gold(Au)). The typical tunnel junction area and contact resistance were ~10-25 $\mu m^2$ and ~1 MΩ respectively for all the MEMED samples. The lower bound of the cross-junction's dimension was chosen according to our photolithography machine resolution limit. The upper bound of the junction's dimension was determined by the aerial density of the shorting defects [17, 19]. Intense experimental efforts were required to optimize the deposition parameters (power, pressure and seed layer material) so that the metal surfaces had < 0.2 nm RMS roughness. Bottom electrode roughness was measured by tapping mode AFM (Digital Instruments Multimode).  Another critical experimental optimization was to minimize the compressive stresses in the deposited films. We observed that the stress release mechanism involved the formation of hillocks that short circuited the tunnel barrier, hence, dramatically reducing the temporal and thermal stability of the device.

## 3. Results and Discussion:

Current-voltage measurements were performed to study the effect of molecular channel(s) on the bare tunnel junction with the exposed side edge(s). Present work focuses on the molecular spintronics devices. For preparing our molecular spin devices based on liftoff-MEMED approach we fabricated tunnel junction test beds with ferromagnetic electrodes or magnetic tunnel junctions (MTJs). MTJs with Co/NiFe bottom electrodes and NiFe top electrodes showed several folds current increase due to molecular channels (Fig. 8).  The Co/NiFe/AlOx/NiFe MTJ is the key tunnel junction configuration which was used in liftoff based MEMSD or liftoff-MEMSD. Although in this paper we only discussed the OMC induced current increase as the signature of molecules' effect, however, liftoff-MEMSDs also produced a number of intriguing observations. For instance, OMCs in a liftoff-MEMSD showed dramatic current suppression and strong effect on the magnetic properties of the MTJ test bed. These observations will be discussed elsewhere. It is noteworthy that Co/NiFe/AlOx/NiFe MTJ configuration resulted from the quest of producing a stable and chemical etch-proof ferromagnetic tunnel junction, not from a premeditated plan. Experimental results about device stability are discussed elsewhere in this paper itself.

OMCs [21] (Fig. 7) produced a clear increase over MTJ's leakage current (Fig. 8a). Similarly, Icosane-1,20 dithiol molecules exhibited the distinct current increase (Fig. 8b). However, the current increase due to molecular channels varied on different MEMEDs (e.g. Fig. 8a vs. Fig. 8c); probably the reason is inconsistent liftoff quality along the MEMED edges or AlOx having higher thickness along the edges (Fig. 6g). To increase the density of molecular conduction channels in liftoff-MEMSD one can further decrease the AlOx thickness (< 1.5 nm). However, this approach will diminish the tunnel junction stability. Relaxing mechanical stresses and electrical bias induced instability keep deteriorating a tunnel junction, which showed high quality current-voltage response in its as made state. To minimize the bias induced instability we used 50-100 mV bias range for current-voltage studies, which is well below the typical tunnel junction's breakdown voltage of 1.5-2.0 V.

On a chip, we simultaneously produced a large number of MEMSDs; scheme of increasing aerial device density was found to affect the yield and ability to conduct complementary control experiments. Generally, we used cross bar (X) device geometry (Fig. 6h) due to following reasons: (a) a set of 36 tunnel junctions could be produced in 1 mm$^2$ area. The dense packing was necessary since we had a small processing area to get promising samples within the sputtering machine. The deposition of high quality and controllable 1.5 nm thick Al was realized through customized shuttering scheme for a small process area on the substrate holder. (b) The quality of plasma oxidation of 1.5 nm Al varied drastically along the radial direction of substrate holder. (c) The chimney form of sputtering gun made deposition quasi-directional; it is noteworthy that the sputtering process is isotropic only for the specific arrangement of sample and sputtering gun. To ensure that interference among devices for the cross bar device scheme did not produce any interference or spurious data MEMED were also produced in "T" shaped isolated tunnel junction (Fig. 6i). Isolated tunnel junction had an advantage that bottom electrode was in mm$^2$ range; localized damage to bottom electrode don't matter. Additionally, the leakage current via T-shape junction was smaller as compared to that of X junction. It is presumably due to the reason that T-junction (X-junction) had one (two) bottom electrode edge(s) in the junction area. Occasionally Bottom electrode edges have unpredictable notches to produce leakage paths or hot spots in the junction area. Longer the bottom electrode edge length higher the leakage current. MEMED with this robust device design yielded similar response from the OMC conduction channels (Fig. 8d). OMC channels along T-junction edges produced distinct increase over the bare tunnel junction current.

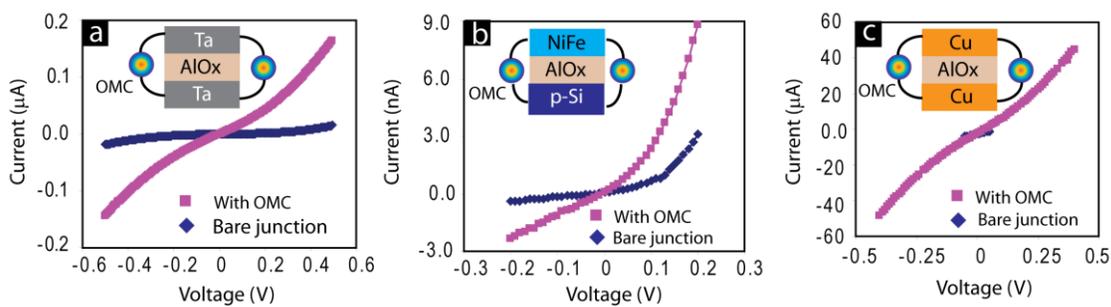

Fig. 9: Current-voltage studies of nonmagnetic tunnel junction based molecular devices: (a) Ta/AlOx/Ta, (b) p-Si/AlOx/NiFe, and (c) Cu/AlOx/Cu tunnel junctions before and after the establishment of molecular conduction channels.

Besides ferromagnetic electrodes, several configurations of the metal electrodes were explored with liftoff-MEMED/MEMSD approach; these attempts are particularly important since most of the molecular devices produced so far utilized gold as the metal electrodes. In one of our liftoff-MEMED Tantalum (Ta) metal was used as the top and bottom electrode. The fabrication protocol was exactly same as shown in figure 6, only instead of ferromagnetic metal Ta was used. The Ta/AlOx/Ta tunnel junction test bed exhibited leakage current in nA range and was 1-2 order smaller than that observed with Co/NiFe/AlOx/ NiFe tunnel junction test bed. This difference was due to the difference in bottom electrode ability to produce an insulator once exposed to oxygen. It is noteworthy that AlOx for the test bed is produced by the plasma oxidation of Al. In the event when surplus oxygen flux is supplied the excess oxygen atoms reaches bottom electrode and oxidize it. When Ta is used as the bottom electrode, in the event of overoxidation it produced TaOx insulator. The effective insulating barrier is expected to be a bilayer AlOx/TaOx insulator. This bilayer insulator produced nA level leakage current via the planar area. Lower leakage current through the test bed tunnel junction is highly desirable to allow the fewer molecular channels to show their effect in current-voltage measurements. On a Ta/TaOx/AlOx//Ta test bed the OMCs increased tunnel junction current by ~50 folds (Fig. 8a). It is noteworthy that upon oxidation unlike Ta, NiFe bottom electrode's oxidation produced iron oxides and elemental Ni after the oxidation step [16]. Thermodynamically, simultaneous oxidation of Ni and Fe is not possible [16]. Therefore, NiFe is unlikely to have a completely oxidized surface. Hence, wherever NiFe is used as the bottom electrode a leakage current is only expected to be due to the AlOx only. Whereas in the case of Ta bottom electrode the leakage current may be due to a bilayer insulator comprising AlOx and TaOx insulator; because during Al oxidation a part of Ta electrode may also get oxidized.

Besides, metallic electrodes p-type Silicon (p-Si) was also used as a bottom electrode in liftoff-MEMED. For the fabrication of this device, the steps for AlOx growth and top metal electrodes were same as shown in figure 6. However, bottom electrode was created by selectively removing SiO$_2$ to expose p-Si in the window of desired dimension. Liftoff step was used for producing exposed side edges of p-Si/AlOx/NiFe junction. Bridging of OMCs along the exposed edges yielded clear increase in charge transport (Fig. 8b). In this case, top electrode is a ferromagnetic metal, i.e. NiFe. One can selectively study the effect of interaction between one magnetic electrode and OMCs. Ferromagnetic electrode can be used as the spin reservoir to supply spin polarized charge into the molecular channel.

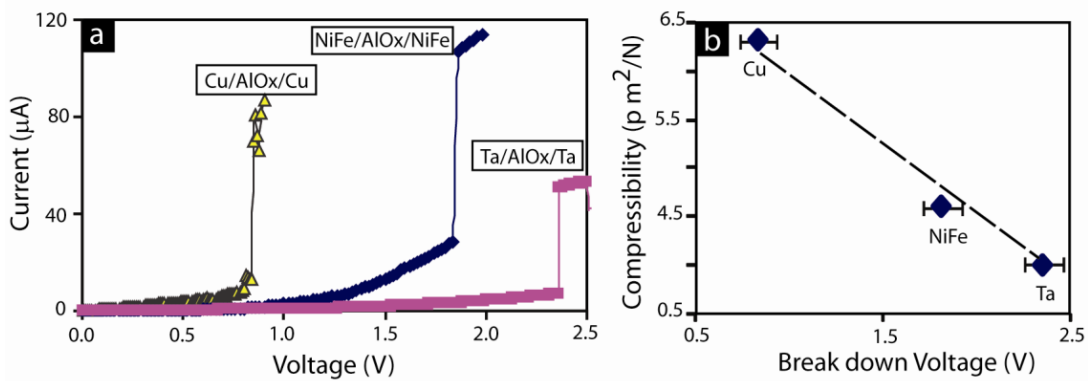

Fig. 10: (a) Current-voltage studies showing breakdown voltage of 2 nm thick AlOx insulator in a tunnel junction with Cu, NiFe and Ta top and bottom electrodes, respectively. (b) Correlation between breakdown voltage of tunnel junctions with Cu, NiFe and Ta with their compressibility.

Si bottom electrode resembled with the Ta bottom electrode because it can also be oxidized to produce an insulator. Therefore, Si/AlOx/NiFe test bed is likely to have double barrier composed of SiOx and AlOx, akin to TaOx/AlOx. As a result a MEMED test bed can be more stable and sturdy if bottom electrode can be oxidized to form an insulator. In the same line we attempted to form liftoff-MEMED using copper (Cu) as the two electrodes. Cu is known to oxidize easily and produce semiconducting oxide. For Cu based liftoff-MEMED fabrication steps were same as shown in figure 6. Like other liftoff MEMEDs the addition of OMCs produced distinctive increase over tunnel junction's leakage current (Fig. 8c). In order to examine the stability of tunnel junction test bed electrical breakdown studies were performed. For this study tunnel junction with NiFe, Cu and Ta bottom metal electrodes were selected and their breakdown voltage was evaluated. It is noteworthy that NiFe metal did not produce an insulator, while Cu and Ta produced oxide layers which behaved as an insulator or semiconductor. For all the cases, tunnel junction were produced by the exactly same procedure (Fig. 6). In the tunnel breakdown study tunnel junction with Cu bottom electrode exhibited the smallest electrical breakdown voltage ~1 V, while tunnel junction test beds with Ta bottom electrode showed the highest electrical break down voltage of ~2.5 V. It is interesting to note that Ta and Cu both produced oxides; hence, just having an oxidizable metal electrode was not the sufficient reason for producing higher breakdown voltage. According to our previous efforts, breakdown of a tunnel barrier can be strongly dependent upon the bottom electrode deformability under the effect of residual mechanical stresses. We observed that a completed tunnel junction possessed compressive stresses. These compressive stresses were found to be responsible for creating nanohillocks out of bottom metal electrodes. These nanohillocks could pierce through the ultrathin insulator above it [17]. The susceptibility of a metal to form nanohillocks can be correlated to its compressibility. We plotted electrical breakdown voltage and the compressibility of bottom metal electrode. Interestingly, electrical breakdown and compressibility varied linearly (Fig. 10b). Hence, while designing a liftoff-MEMED it is imperative to study the effect of the bottom electrode on the tunnel junction stability. In addition, many metals are not compatible with the protocol used for bridging molecular channels. For instance, Cu was susceptible for the etching during prolong electrochemical molecule attachment step, while Ta and NiFe metals remained undamaged by the same step.

To ascertain that only molecule enhanced the charge transport at the tunnel junction test beds several control experiments were performed [11]. In the simplest control experiment, the insulator thickness was set to be more than the molecule's physical length. Tunnel junction with thicker tunnel barrier (> molecule length) did not show any current increase. In another control experiment molecule with only single thiol functional group were utilized. Since these molecules can not bridge the insulator gap hence no effect on the transport of the bare tunnel junction was expected. Tunnel junction subjected to molecules with monothiol functional group indeed did not affect the tunnel junction transport. These two control experiments also suggested that a stable tunnel junction with exposed side edges will only show current enhancement when molecular channels are bridged across the tunnel barriers, not due to mere random chemisorption.

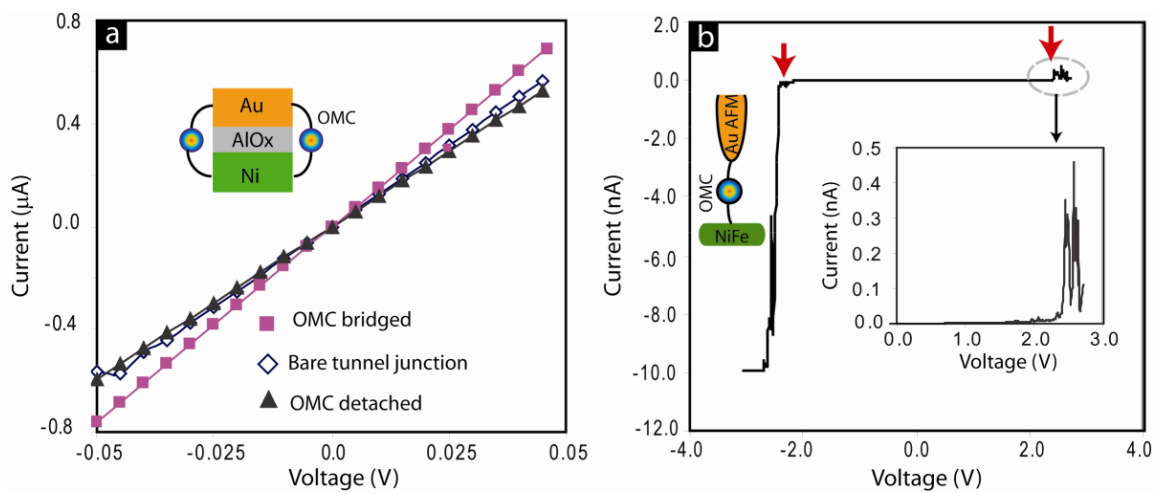

Fig. 11: (a) Reversible effect of OMC on Ni/AlOx/Au tunnel junction transport and (b) transport through single OMC studied by conducting probe AFM. Note in panel (a) and (b) two metal leads connected to OMC (s) are analogous to each other.

In a crucial experiment, molecular channels' effect was reversed to retrieve the charge transport characteristics of a tunnel junction test bed. For this study Ni/AlOx/Au tunnel junctions and OMCs were utilized (Fig. 11a). OMCs chemically bonded to Ni and Au electrodes to establish the bridges for additional charge transport; OMC's thiol functionalized tethers (Fig. 7b) produced Ni-S and Au-S linkage to complete molecular bridging. The Ni-S and Au-S bonds have significantly dissimilar bond energy. After obtaining the increase in tunnel junction's charge transport due to OMCs, Au-S bonds were selectively broken. To do so the sample was submerged in the 5 mM dodecane dithiol solution. Mass action of the dodecane dithiol molecules selectively broke the Au-S bonds to undo the effect of molecular channels (Fig. 11a). On the same sample OMCs attachment and detachment steps were repeated twice. We also attempted to remove the OMCs by alternative method. Under this alternative approach plasma oxidation was used for burning OMCs and to retrieve the bare tunnel junction transport characteristics. However, plasma oxidation caused tunnel junction failure, leading to a resistor type charge transport profile.

MEMEDs and MEMSDs utilized several thousands of OMCs. We investigated the correlation between the charge transport through the single OMC channel and simultaneous charge transport through thousands of OMCs. For the study of charge transport through the single OMC, a self assembled OMC monolayer on NiFe flat film was studied by the conducting probe atomic force microscope (CPAFM) probe. CPAFM probe contacted one OMC at a time as per arrangement shown in the schematic in figure 11b. The initial current-voltage response appeared with a nonconducting gap and then sudden increase in current past 2 V; within nonconducting gap current magnitude was of the order of ~0.1 pA. This current-voltage profile indicates the presence of ~ 2 V charging barrier height; electrons on the metallic electrode has to overcome this charging barrier to reach the molecular energy levels arising from the organometallic center cube of an OMC (Fig. 7). Presence of charging barrier suggested that the mechanism of charge transport through an OMC is coherent tunneling. The dissimilar current magnitudes for the opposite bias regimes can be attributed to the difference in electrical contacts with the studied OMC. We also observed that after repeating current-voltage studies the width of non-conducting regime decreased. We attributed these changes to the changing molecular configuration, the inter-electrode separation, molecular charge state etc.

For a close comparison of single OMC transport data we considered Ni/AlOx/Au based MEMED (Fig. 11a) as its configuration conforms with the two metal electrodes used in CPAFM study (Fig. 11b). OMCs attached to thin film electrodes of a tunnel junction mainly showed a tunneling type charge transport (Fig. 11a). Tunneling type response is clearly observable for the high bias transport data (Fig. 9 a and c). Since the electrical breakdown of test bed tunnel junction was ~1.6 V, hence we were unable to capture charge transport characteristics for higher bias range in 2.5-3 V range. Recording current-voltage data up to 3.0 V range could enable the observation of the onset increase in molecular transport, similar to the one observed in current-voltage response of single OMC (Fig. 11b).

### 3.1 Analysis of MEMED/MEMSD Current-Voltage Data:
Analyzing current-voltage studies with tunneling models can produce a number of useful system properties such as barrier thickness and barrier height about the tunnel junction test bed [23, 24]. More importantly, once a tunnel junction is transformed into a MEMED or MEMSD then analyzing the transport data with tunneling model or other appropriate charge transport models can give direct understanding about the charge transport mechanism through the molecules [11]. However, for accurate modeling and analysis experimental current-voltage data should be recorded at least in medium voltage range or ± 0.4 V range. We observed that although tunnel junctions showed 1-2 V breakdown voltage, but doing current-voltage studies in medium voltage range occasionally destabilized the tunnel junction. Our experimental studies strongly suggested that a tunnel barrier failure is the combinatorial effect of electrical and mechanical stresses. In addition, an insulator performed very differently on different bottom electrode materials (Fig. 10a). In order to minimize the destabilization of an insulating barrier due to high bias application we generally performed low bias (± 0.05 to 0.1V) current-voltage studies before and after transforming a tunnel junction test bed into a molecular device. This strategy was especially helpful in prolonging the device life of tunnel junction with Co, Cu and gold metal electrodes. Test bed's longevity is crucial since a degrading tunnel junction can mimic the molecule effect and hence will produce inconclusive molecule response. Low-bias current-voltage studies were also helpful in reducing the screening time for the batches of hundreds of tunnel junction test beds; since many tunnel junctions also kept degrading with time due to relaxing mechanical stresses hence it was necessary to complete molecular device fabrication and all the characterizations during the stable stage of a tunnel junction. For extracting barrier properties, the modeling protocol required current-voltage data in ± 0.3-0.5 V range, while our experimental data was in ± 0.05 to 0.1V range. Occasionally the high bias current-voltage data for transport analysis was extrapolated from low bias current-voltage data, using a 3rd order polynomial fit. It is noteworthy that polynomial fit are customarily utilized in the barrier property calculation [25].

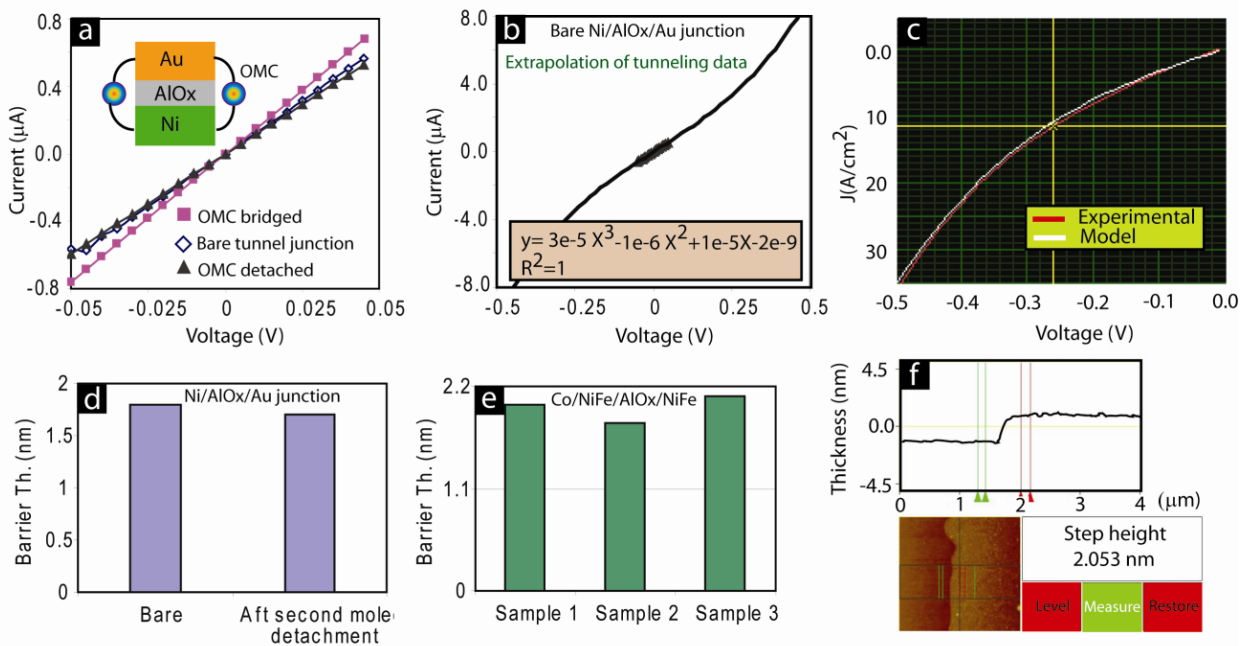

Fig. 12: Supporting experiments: (a) Reversible attaching and detaching molecular clusters produced reversible changes on tunnel junction current. (b) Low bias current-voltage data of bare Ni/AlOx/Au was extrapolated using 3$^{rd}$ order polynomial for modeling. (c) Such extrapolated data were compared with Simmons tunneling model to extract the barrier height and barrier thickness (red line-Simmons model, white line-extrapolated experimental data). (d) Calculated AlOx barrier thickness of the Ni/AlOx/Au before attaching OMC and after reversing the OMC effect. (e) Calculated thickness of AlOx with different Co/NiFe/AlOx/NiFe samples produced over eight months. (f) Experimentally measured AlOx barrier thickness around the time when samples shown in panels (d-e) were fabricated.

To show the utility of extrapolated data in the context of MEMED/MEMSDs let discuss this strategy for the sample on which we proved the successful bridging of molecular channels. This sample was the Ni/AlOx/Au test bed on which we successfully attach and detach the OMCs. After OMCs detachment Ni/AlOx/Au retrieved the transport characteristics of the bare tunnel junction (Fig. 12a). A 3rd order polynomial fit of ± 0.05 V range experimental data and then extrapolating polynomial equation to ± 0.5 V range produced the high bias transport data (Fig.12b). For the present polynomial fit and extrapolation, bare Ni/AlOx/Au tunnel junction data was utilized. After converting the current data into current density data we compared the extrapolated experimental data with the current density-voltage data generated by the Simmons model for medium bias range [26]. Typically, a good agreement was observed between extrapolated data and model produced data (Fig.12c). Subsequently the barrier thickness and barrier height parameters for which model produced the best match to experimental current-voltage data were recorded as the representation of actual barrier properties.

For the consistency check, we performed the same extrapolation and modeling approach to calculate the barrier properties for the Ni/AlOx/Au tunnel junction after undoing the molecule effect. The calculated barrier thicknesses for the bare tunnel junction and after undoing the molecule effect were in excellent agreement (Fig. 12d). Similar extrapolation and modeling approach was applied for a number of magnetic tunnel junctions (MTJs) which were utilized for producing MEMSDs. The calculated barrier thickness for three MTJs over a period of eight months is shown in figure 12e. All the calculated thicknesses were close to 2 nm. For a cross check calculated thicknesses were compared with the physical thickness of AlOx tunnel barrier by the atomic force microscope (AFM). For the physical measurement, ~ 2 nm AlOx deposition utilized exactly the same experimental protocol as used for growing ~ 2 nm AlOx in tunnel junction test beds on which current-voltage data was collected and used for extrapolation by polynomial fit. The calculated thicknesses were in excellent agreement with the physically measured AlOx thickness by AFM (Fig. 12f).

AFM based studies are highly recommended as a monitoring tool to ensure that tunnel junction test beds have required insulator thickness. The quality of the sputtered Al is transient in nature since Al sputtering target keep changing its surface profile; changed target surface directly affect the deposition rate and AlOx quality. Hence, one needs periodic monitoring of the Al thickness. Moreover, any alteration in the device fabrication scheme can also affect the leakage current, which will in turn affect the soundness of calculated tunnel barrier properties through modeling. As a bottom line, it is extremely important to keep AlOx thickness in check by periodic AFM or direct physical measurements, and not to rely on calculated insulator thickness.

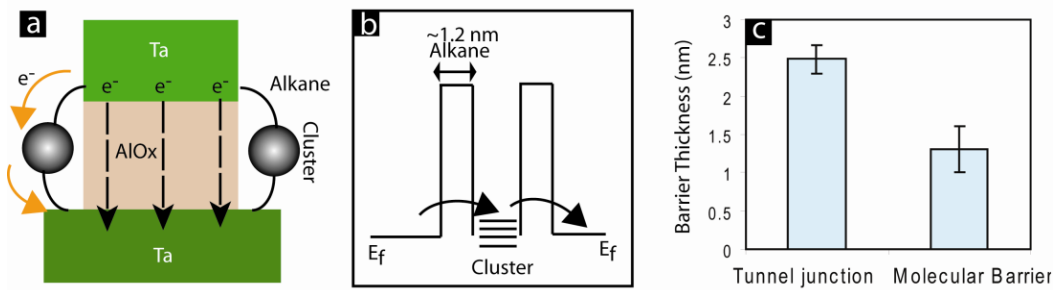

Fig. 13: Mechanism of charge transport through molecule. (a) Schematic of the cross section of MEMED for which as recorded high bias (±0.5 V) electron transport through Ta/AlOx/Ta tunnel junction's planar area and the molecular channels were used for modeling. (b) Charge transport via molecular channels was akin to charge transport via double barrier tunnel junction. (c) Effective barrier thickness after molecule attachment was comparable with alkane tether length.

To study the effect of molecules on the properties of a test bed the transport data were analyzed with the Simmons tunneling model. *For Simmons modeling as received high bias data on Ta/AlOx/Ta tunnel junctions before and after OMC attachment, was utilized* [11]. Properties of the OMC bridges were extracted from the molecular transport data (total current after molecular bridging minus bare tunnel junction's leakage current) (Fig.13a). The effective tunnel junction area was $2.5 \times 10^{-7}$ cm$^2$. The effective molecular area was estimated by assuming that effective 10 μm long side edges of a tunnel junction test bed are covered by the 1 nm wide OMCs; the effective molecular area was 10 μm x 1nm=.$10^{-10}$ cm$^2$. The effective molecular barrier thickness was found to be ~1.2 nm; whereas, overall molecule length was ~ 3 nm. This analysis suggested that alkane tethers controlled the OMC's charge transport. Alkane tethers helped attach the OMCs with the metallic electrodes (Fig. 13b). Before the integration of OMCs barrier thickness was ~2.2 nm for Ta/AlOx/Ta tunnel junction test bed (Fig. 13c). The barrier height was also obtained from the transport modeling. The calculated barrier height was found to be ~0.7 V [11]. It is noteworthy that from our CPAFM study the barrier height was ~2.0 V (Fig. 11b). This difference in the barrier height is presumably due to the nature of chemical bonding with OMC and the shape of metal electrode as well. Additionally, molecular transport was also found to be thermally activated. The temperature dependent charge transport necessitates the extensive current temperature studies at different temperatures. Application of pure tunneling model can only give limited or incomplete understanding of charge transport mechanism. However, our tunnel junction test beds tended to fail during heating and cooling cycle [17], due to which we were unable to perform extensive temperature dependent charge transport studies.

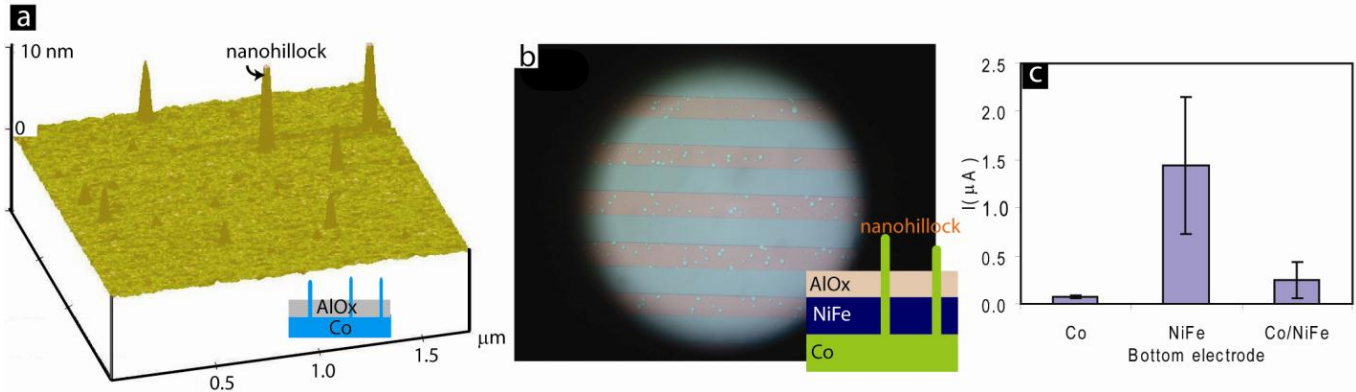

Fig. 14: Bottom electrode instability: (a) AFM study of Co(10 nm)/AlOx (2 nm) showing evolution of hillocks due to compressive stress relaxation. (b) Co (12-14 nm) under AlOx and NiFe (5 nm) produced hillocks; acid etching resulted in hillocks removal and the creation of holes shown in optical photograph, (c) Leakage current at 100 mV through as made tunnel junction with Co, NiFe, and Co/NiFe bottom electrodes.

**3.2 MEMED stability and basis of conceiving the MEMSD configuration:**

Liftoff based MEMEDs were the basis of producing MEMSDs, the central theme of this work. Realization of a stable MEMSD required (a) fabrication of a stable ultrathin insulator on the specific bottom metal electrodes and (b) selection of spin polarized metallic electrodes, which can withstand molecular solution and electrochemical step required to bridge the molecular channels. After all the optimization efforts the final magnetic tunnel junction came out to be quite different than what we planned. Moreover this configuration exhibited a number of highly intriguing phenomenons on the molecular spin devices for the first time. For producing MEMSD, initially we prepared Co/AlOx/Co and NiFe/AlOx/NiFe tunnel junctions with the exposed side edges using liftoff approach. Typically, 10-12 nm thicknesses of individual electrodes were utilized; 10-12 nm ferromagnetic electrode thickness was amenable for clean liftoff and also enabled the robust electrical connection during transport studies by micromanipulators. Interestingly, AlOx insulator behaved quite differently on different metallic electrodes. For the same AlOx thickness tunnel junction with Co bottom electrode showed ~1 order smaller leakage current than the current observed with NiFe bottom electrode. This observation created a temptation to use Co metal electrode. However, tunnel junctions with Co metal electrode showed multiple issues. Tunnel junctions with Co metal electrode showed time dependent tunnel barrier failure. From Co bottom electrode nanohillocks evolved to pierce out of AlOx insulator (Fig. 14a). Nanohillocks evolved with time as process induced compressive stresses relaxed [17]. These nanohillocks poked through AlOx tunnel barrier to create short circuit. On the other hand no nanohillock was observed when Co was replaced with the NiFe bottom electrode. Additionally, NiFe is much more oxidation resistant [16] and hence replacing Co with NiFe make tunnel junction oxidation resistant.

The Co metal was not only prone to oxidation and nanohillocking but also got damaged by our molecular solution and electrochemical molecule attachment protocol. Due to these unfavorable attributes, Co metal electrode could not be used for MEMSD fabrication. On the other hand, NiFe is free from both of these disadvantages, and hence was suitable for the MEMSD fabrication. However, leakage current for NiFe bottom electrode was ~10 fold higher than that observed with Co bottom electrode for the same AlOx thickness. It was also found that Co(5-7nm)/NiFe (3-5 nm) bilayer worked optimally as an bottom electrode. A Co/NiFe bottom electrode did not show significant evolution of deleterious nanohillocks unless Co thickness was >10 nm. Significant nanohillocking occurred for 12-14 nm thick Co even under the NiFe and AlOx. For a Co(12-14 nm)/NiFe (5nm) bilayer during electrochemical molecule treatment [11] regions at the nanohillocks sites got etched away. Etching of nanohillocks left easily perceivable holes in the bottom electrode (Fig. 14b). The tunnel junction with Co(5-7 nm)/NiFe (3-5nm) exhibited leakage current significantly smaller than that observed with NiFe bottom electrode (Fig. 14c). Cumulatively, MEMSD devices had Co/NiFe bottom electrodes and NiFe top electrode. These top and bottom electrodes possessed significantly different magnetization and spin wave excitation properties. Difference in magnetic properties of the two electrodes was a crucial prerequisite to observe intriguing properties with molecular spin devices. For instance, a molecular spintronics device utilizing a break-junction with two ferromagnetic electrodes of different saturation magnetization exhibited anomalous Kondo splitting [3]. In summary, we initially intended to use single layer Co and NiFe metal electrodes in MEMSD; however, due to experimental difficulties we settled for bilayer bottom electrode. Further improvements were also observed when tantalum (Ta) seed layer was deposited before the deposition of bottom electrode.

A liftoff based MEMSD is prone to have tall notches along the tunnel barrier edges (Fig. 6g). To mitigate this possibility one need to optimize the second photolithography step and thin film deposition conditions. Other more practical approach may be to introduce a brief etching step after the fabrication of a tunnel junction, with exposed side edges from the liftoff method. One needs to etch ~50 nm width of insulator starting from the exposed edge regions to get rid of edge notches. Removal of notches from the edge region will ensure that the minimum distance between the two metallic electrodes along the exposed edge region is equal to the insulator film thickness and hence increasing the MEMED/MEMSD yield.

## 4. Summary:

A liftoff based multilayer edge molecular device approach enabled the transformation of a magnetic tunnel junction into a molecular electronics or spintronics device. Multilayer edge molecular spintronics devices (MEMSD) can be produced with a variety of electrode configurations. The deposition of ultra thin insulator with low leakage current is the critical element in producing MEMSD. MEMSD can be used as a test bed for a variety of molecules, including single molecular magnets. Fabrication of a stable and long living MEMED or MEMSD necessitate the careful consideration of destabilizing factors appearing during and after the tunnel junction fabrication. Considering the stress induced device failure and etching induced damage from the molecular solution to Co metal electrode, we started using Co/NiFe bilayer bottom metal electrode. Specifically, we utilized Co(5-7 nm)/NiFe(3-5 nm)/AlOx/NiFe(10-12 nm) magnetic tunnel junction configuration for fabricating MEMSDs. Magnetic studies before and after molecule attachment for this configuration is discussed elsewhere [27].

**Acknowledgments:**

PT thanks Prof. Bruce J. Hinds and the Department of Chemical and Materials Engineering, University of Kentucky to enable his PhD research work presented in this manuscript. He also thanks D.F Li and S. M. Holmes for providing molecules used in this work.